\documentclass[%
 reprint,
superscriptaddress,
 amsmath,amssymb,
 aps,
pre,
]{revtex4-2}

\usepackage{graphicx}
\usepackage{dcolumn}
\usepackage{bm}
\usepackage[colorlinks=true]{hyperref}
\usepackage{mwe}
\usepackage{color}
\usepackage[dvipsnames]{xcolor}

\newcommand{\kB}{k_\mathrm{B}}
\newcommand{\DmX}{\Delta\mu^X}
\newcommand{\DmY}{\Delta\mu^Y}
\newcommand{\fof}{\mathrm{F}_{\rm o}\!-\!\mathrm{F}_1}

\newcommand{\sfu}{\affiliation{%
 Department of Physics, Simon Fraser University, Burnaby, BC, V5A 1S6 Canada
}}

\newcommand{\erlangen}{\affiliation{%
 PULS Group, Department of Physics, FAU Erlangen-Nürnberg, IZNF, Erlangen, Germany
}}

\begin{document}

\preprint{APS/123-QED}

\title{Unlocking the potential of information flow: Maximizing free-energy transduction in a model of an autonomous rotary molecular motor}

\author{Mathis Grelier}
\email{mathis.grelier@fau.de}
\sfu
\erlangen

\author{David A.\ Sivak}
\email{dsivak@sfu.ca}
\sfu

\author{Jannik Ehrich}
\email{jehrich@sfu.ca}
\sfu

\date{\today}

\begin{abstract}
Molecular motors fulfill critical functions within all living beings. Understanding their underlying working principles is therefore of great interest. Here we develop a simple model inspired by the two-component biomolecular motor Fo-F1 ATP synthase. We analyze its energetics and characterize information flows between the machine's components. At maximum output power we find that information transduction plays a minor role for free-energy transduction. However, when the two components are coupled to different environments (e.g., when in contact with heat baths at different temperatures), we show that information flow becomes a resource worth exploiting to maximize free-energy transduction. Our findings suggest that real-world powerful and efficient information engines could be found in machines whose components are subjected to fluctuations of different strength, since in this situation the benefit gained from using information for work extraction can outweigh the costs of information generation.
\end{abstract}

                              
\maketitle                              

\section{Introduction}
In the last two hundred years, thermodynamics has traversed a grand journey, from its inception with the desire to describe the workings and performance limits of steam engines, to recent advances in stochastic thermodynamics~\cite{Jarzynski2011_Equalities,Seifert2012_Stochastic,VandenBroeck2015_Ensemble,Peliti2021_book}. This theory enables understanding of the energetics~\cite{Brown2020_Theory} of nanoscale engines such as molecular motors~\cite{Chowdhury2013_Stochastic,Kolomeisky2013_Motor}. Applying the theory to models of real-world molecular machines gives insight into their inner workings and can lead to useful design principles that may be applied in the development of artificial engines.

An important subclass of molecular machines consists of two strongly coupled components, such as $\mathrm{F}_{\rm o}\!-\!\mathrm{F}_1$ ATP synthase~\cite{Junge2015_ATP_Synthase}, which turns one form of chemical energy into another by converting a proton gradient into the production of ATP. While the free-energy transduction in such two-component motors is mechanical in the sense that different components move relative to each other, the exact mechanism leading to optimal performance remains opaque. Specifically, there are two distinct pathways available for free-energy transduction~\cite{Ehrich2022_Energy}: (1) an upstream system generates an \emph{energy flow} that directly drives the downstream system; and (2) an upstream system generates an \emph{information flow}~\cite{Horowitz2014_Thermodynamics} that is exploited to rectify thermal fluctuations of the downstream system into output power~\cite{Reimann2002_Brownian}.

The role of information flow in bipartite systems~\cite{Barato2013_Information-theoretic,Hartich2014_Transfer_entropy,Diana2014_Mutual} has recently been reviewed in Ref.~\cite{Ehrich2022_Energy}. Notably, in coupled bipartite work transducers, information flow and energy flow are combined to form a rate of transduced free energy~\cite{Large2021_Free-energy}: Their sum~\cite{Barato2017_Thermodynamic_cost} equals the \emph{transduced capacity}~\cite{Lathouwers2022_Internal}, a kind of bottleneck that simultaneously bounds input and output power and allows definition of subsystem efficiencies~\cite{Leighton2022_Inferring}.

The cellular environment of molecular machines can also feature strong nonequilibrium fluctuations~\cite{Mizuno2007_Nonequilibrium,Gallet2009_Power} generated, e.g., by active media~\cite{Ramaswamy2010_Mechanics, Marchetti2013_Hydrodynamics,Elgeti2015_Physics,Bechinger2016_Active}. The thermodynamics~\cite{Speck2016_Stochastic,Fodor2016_How,Mandal2017_Entropy,Pietzonka2018_Entropy,Dabelow2019_Irreversibility,Caprini2019_Entropy,Dabelow2021_Irreversibility,Datta2022_Second_law} of such active media and ratchet effects~\cite{Sokolov2010_Swimming, Leonardo2010_Bacterial, Reichhardt2017_Ratchet, Vizsnyiczai2017_Light,Pietzonka2019_Autonomous,Fodor2021_Active} within them have recently gained attention. These active fluctuations have the potential to speed up molecular transport motors~\cite{Ariga2021_Noise-Induced} and enzymatic catalysis~\cite{Tripathi2022_Acceleration}. It has also recently been demonstrated that active noise increases the output power of information engines~\cite{paneru22,malgaretti22,Saha2022_Information} and enables extraction of more work than is needed to perform feedback control on it~\cite{Ehrich2022_Energetic}, leading to a useful work-extraction machine.

The dynamics of coupled work transducers~\cite{Golubeva2012_Efficiency,Fogedby2017_Minimal_model,Sune2019,Lathouwers2020_Nonequilibrium,Lathouwers2022_Internal} have been described with models having continuous state spaces. Reference~\cite{Amano2022_Insights} analyzes energy and information flows in a synthetic molecular motor, and Refs.~\cite{Horowitz2013_Imitating,Schmitt2015_Molecular} compare the relative merits of chemical-driving and information-ratchet mechanisms. 

In this paper we investigate the interplay between energy and information flows, elucidating the circumstances under which either of them maximizes free-energy transduction. To this end we consider a minimal two-component work transducer with a discrete state space, simplifying an earlier model with continuous degrees of freedom~\cite{Lathouwers2020_Nonequilibrium,Lathouwers2022_Internal}, and optimize its energy landscape.

In this paper we: (1) introduce a discrete model inspired by the rotary molecular motor $\mathrm{F}_{\rm o}\!-\!\mathrm{F}_1$ ATP synthase that sustains energy and information flows between the motor's components; (2) optimize the model's parameters to reach optimal output power; and (3) analyze the motor when the two components are at different temperatures, mimicking some aspects of nonequilibrium active noise that only acts on one motor component and thereby illustrating that information flow is a valuable resource when additional fluctuations are present.

\section{Discrete work transducer model}
Here we introduce the model and describe its dynamics and thermodynamics.

\subsection{Model dynamics}
\begin{figure}[tb]
    \includegraphics[width=\linewidth]{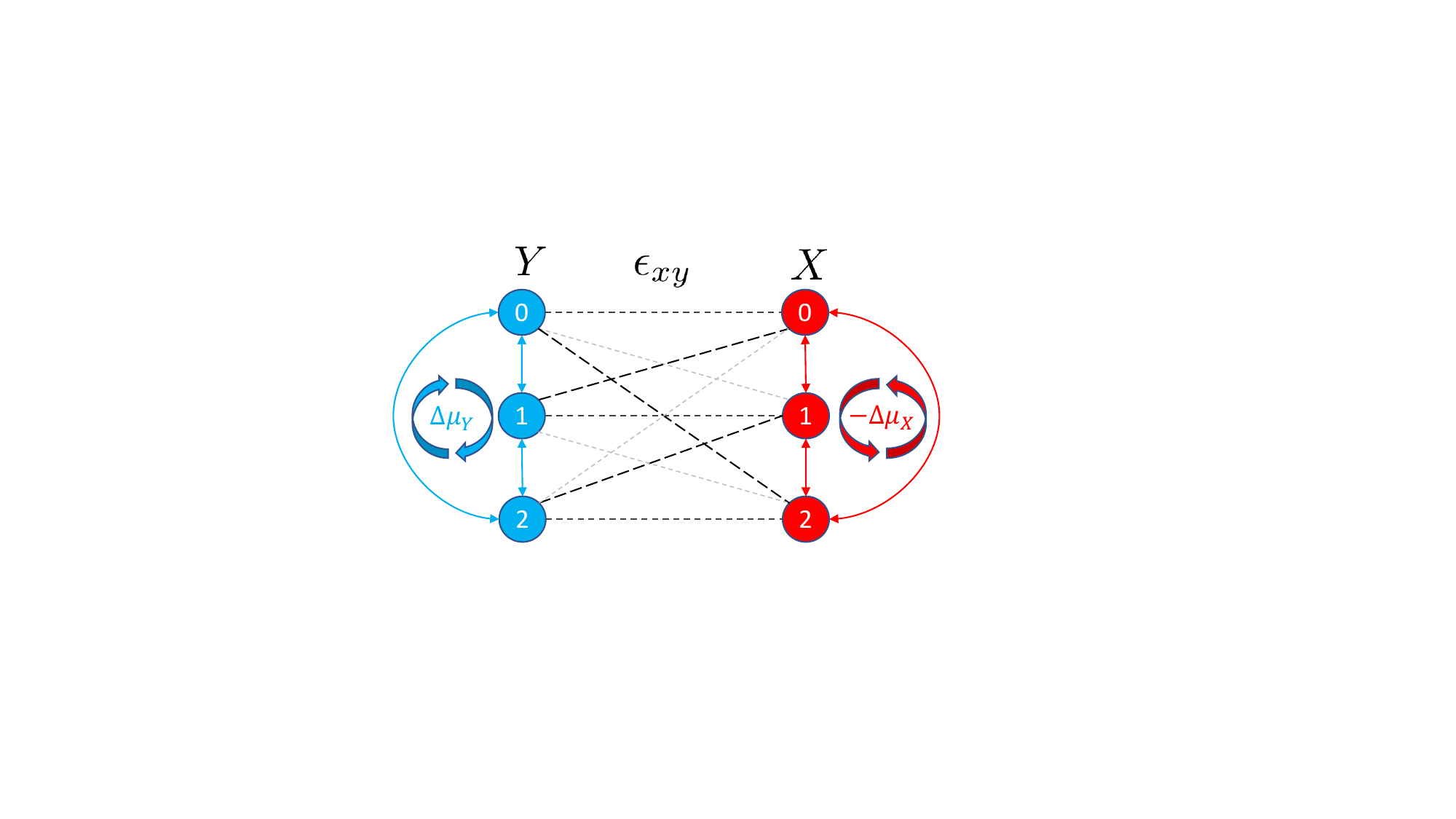}
    \caption{State graph of the system. The $Y$-subsystem (blue) and $X$-subsystem (red) each have 3 states. Thin solid arrows represent possible transitions in each subsystem. Thick cyclic arrows represent the direction of each subsystem's driving force. Dashed lines indicate the internal energy in each state of the joint system, from low (light gray) to high (black).}
    \label{fig:states_graph}
\end{figure}

Our model is inspired by $\fof$ ATP synthase, which consists of two rotating subsystems: $Y$ represents the $\mathrm{F}_{\rm o}$ subunit, while $X$ is the $\mathrm{F}_1$ subunit. Real molecular motors operate in continuous state spaces with energetic minima corresponding to mesostates. Here, we use a discrete state space, capturing the most relevant dynamics. Each subsystem has three states ($0,1,2$), and a transition between two states corresponds to a rotation of $120^{\circ}$. Figure~\ref{fig:states_graph} shows a graph of the system states. The two coupled subsystems are each in contact with a heat bath at temperature $T$ and are driven in opposite directions by opposing chemical-potential gradients: a constant positive gradient $\Delta\mu^Y >0$ drives $Y$ clockwise, while $X$ is driven counter-clockwise by a constant negative gradient $\Delta\mu^X < 0$.

The internal energy of each of the 9 states of the joint system is encoded in a $3\times 3$ matrix $\epsilon_{xy}$, encompassing potential energies specific to each subsystem, as well as coupling energy. We assume that the system is Markovian, and bipartite such that only one subsystem changes its state at a time, constraining the rates $R^{xx'}_{yy'}$ for the transition from $(x',y')$ to $(x,y)$ to
\begin{align}
R^{xx'}_{yy'} = \left\{
    \begin{array}{ll}
        R^{xx'}_{y} & \mbox{if}\; x\ne x', y=y' \\
        R^{x}_{yy'} & \mbox{if}\; x=x', y\ne y'\\
        0 & \mbox{if}\; x\ne x', y\ne y'\\
        -\sum\limits_{x\ne x'}R^{xx'}_{y}-\sum\limits_{y\ne y'}R^{x}_{yy'} & \mbox{if}\; x=x',  y=y'
    \end{array}
\right.\,,  \label{eq:bipartite_assumption}
\end{align}
where the last line encodes conservation of probability. The system is autonomous: the transition rates do not vary in time. Hence we only consider the steady-state distribution of system states, where the joint system's dynamics are described by the stationary master equation
\begin{align}
    \sum_{x',y'} J_{xx'}^{yy'} = 0\,, \label{eq:master_eq}
\end{align}
for probability current $J_{xx'}^{yy'} := R^{xx'}_{yy'} \, p_{x'y'} - R^{x'x}_{y'y} \, p_{xy}$ from state $(x',y')$ to state $(x,y)$, and probability $p_{xy}$ to find the joint system in state $(x,y)$. Along a given transition, the forward and reverse rates are related by the local detailed-balance condition~\cite{Bergmann1955_New,VandenBroeck2015_Ensemble,Seifert2019_From,Maes2021_Local},
\begin{subequations}
\begin{align} 
    \frac{R^{xx'}_{yy'}}{R^{x'x}_{y'y}} &= \exp\left(-\frac{q_{xx'}^{yy'}}{\kB T}\right) \label{generalized balance heat} \\
    &= \exp\left(\frac{\epsilon_{x'y'}-\epsilon_{xy}-\Delta\mu^{xx'}_{yy'}}{\kB T}\right)\,,\label{eq:generalized_det_balance}
\end{align}
\end{subequations}
where $q_{xx'}^{yy'}$ is the heat flowing into the system during transition $(x',y') \to (x,y)$, $\Delta\mu^{xx'}_{yy'}$ is the free-energy change in the chemical reservoir associated with the transition, and $\kB$ is Boltzmann's constant.

Rescaling energies by $\kB T$, substituting the bipartite assumption~\eqref{eq:bipartite_assumption} and the local detailed-balance condition~\eqref{eq:generalized_det_balance}, we explicitly write the transition rates as
\begin{subequations}
\begin{align}
     R^{xx'}_{y} & = \Gamma^{X} \exp\left(\frac{\epsilon_{x'y} - \epsilon_{xy} \pm \Delta\mu^X}{2}\right)
     \label{eq:transition_rates_x}\\
     R^x_{yy'} &= \Gamma^{Y} \exp\left(\frac{\epsilon_{xy'} - \epsilon_{xy} \pm \Delta\mu^Y}{2}\right)\,,\label{eq:transition_rates_y}
\end{align}
\end{subequations}
where $\DmX$ and $\DmY$ are the chemical driving forces on the system, i.e., the negative free-energy changes in the respective reservoirs when $X$ or $Y$ advance their state. The plus sign holds for a forward transition, i.e., when $x = (x'+1) \mod 3$, while the minus sign holds for a backward transition, i.e., $x = (x'-1) \mod 3$, and similarly for $Y$. The kinetic prefactors $\Gamma_X$ and $\Gamma_Y$ represent the bare transition rates, in the absence of energy differences between states. For simplicity, we use equal bare rates and rescale time such that $\Gamma^X = \Gamma^Y = 1$. We also assume that energy changes affect the transition rates in a symmetric way. The effect of alternative choices is discussed in Appendix~\ref{app:splitting_factors}.

\subsection{Thermodynamics}
The bipartite assumption~\eqref{eq:bipartite_assumption} permits identification of the individual probability currents $J^{xx'}_{y}$ and $J^x_{yy'}$ associated with the net transition rates between adjacent states in each subsystem,
\begin{subequations}
\begin{align}
    J^{xx'}_{y} &:= R^{xx'}_ {y}p_{x'y} - R^{x'x}_{y}p_{xy} \label{probability current_x}\\
    J^{x}_{yy'} &:= R^{x}_ {yy'}p_{xy'} - R^{x}_{y'y}p_{xy}\,. \label{probability current_y}
\end{align}
\end{subequations}
These definitions in turn decompose the various flows of energy and information into contributions due to the individual dynamics of each subsystem~\cite{Ehrich2022_Energy}.

Each state transition is accompanied by an exchange of chemical energy between the joint system and one of the two chemical reservoirs. Thus, we identify the average rates of work done on the individual subsystems,
\begin{subequations}
\begin{align}
     \dot W^X & = \sum_{y} \left(J^{10}_{y} + J^{21}_{y} + J^{02}_{y}\right) \Delta\mu^X \label{eq:work_flows_x} \\
     \dot W^Y & = \sum_{x} \left(J_{10}^{x} + J_{21}^{x} + J_{02}^{x}\right) \Delta\mu^Y\,. \label{eq:work_flows_y}
\end{align}
\end{subequations}
By convention, positive work indicates that work is done on the system. Thus to transduce energy from $Y$ to $X$, we require $\dot W^X < 0$ and $\dot W^Y >0$, meaning that $Y$ moves in the direction of its chemical-potential gradient, whereas $X$ is driven against its own gradient. This allows us to define an efficiency as the ratio
\begin{align}
    \eta := \frac{-\dot W^X}{\dot W^Y} \label{eq:efficiency}
\end{align}
of output power to input power.

Beyond energy exchanged with the chemical reservoirs, there is also energy exchanged with the heat reservoir. The average rates of heat flowing into each subsystem are~\cite{VandenBroeck2015_Ensemble,Seifert2012_Stochastic,Ehrich2022_Energy}
\begin{subequations}\label{eq:heat_flows}
\begin{align}
     \dot Q^X & = -\kB T\sum\limits_{x\ge x',y} J^{xx'}_{y} \ln\frac{R^{xx'}_{y}}{R^{x'x}_{y}} \label{eq:heat_flows_x} \\
     \dot Q^Y & = -\kB T\sum\limits_{x,y\ge y'} J^{x}_{y'y} \ln\frac{R^{x}_{yy'}}{R^{x}_{y'y}} \,. \label{eq:heat_flows_y}
\end{align}
\end{subequations}
When one subsystem changes its state, it also changes the potential energy $\epsilon_{xy}$ and hence does work on the other subsystem. We therefore identify the average rates of transduced work~\cite{Large2021_Free-energy,Ehrich2022_Energy} as
\begin{subequations}
\begin{align}
    \dot W^{X\rightarrow Y} &:=\sum_{x \geq x',y} J^{xx'}_{y} (\epsilon_{xy} - \epsilon_{x'y})\label{eq:internal_energy_flow_x}\\
    \dot W^{Y\rightarrow X} &:=\sum_{x,y\geq y'} J_{yy'}^{x} (\epsilon_{xy} - \epsilon_{xy'})\label{eq:internal_energy_flow_y}\,.
\end{align}
\end{subequations}
The system is at steady state and therefore the average internal energy is constant,
\begin{align}
    0 = \mathrm{d}_t E = \dot W^{X \to Y} + \dot W^{Y \to X}\,, 
\end{align}
where the second equality requires the bipartite assumption~\eqref{eq:bipartite_assumption}. Hence $\dot W^{X \to Y} = - \dot W^{Y \to X}$. Together with the local detailed-balance condition~\eqref{eq:generalized_det_balance}, the definitions in Eqs.~(\ref{eq:work_flows_x},\ref{eq:work_flows_y}), (\ref{eq:heat_flows_x},\ref{eq:heat_flows_y}), and (\ref{eq:internal_energy_flow_x},\ref{eq:internal_energy_flow_y}) fulfill subsystem-specific first laws:
\begin{subequations}\label{eq:first_law}
\begin{align}
    -\dot W^{Y \to X} &= \dot W^X + \dot Q^X \label{eq:first_law_x} \\
    \dot W^{Y \to X} &= \dot W^Y + \dot Q^Y\,. \label{eq:first_law_y}
\end{align}
\end{subequations}
Similar to the splitting of the energetic quantities above, the average entropy production rate can be split into nonnegative subsystem-specific components~\cite{Hartich2014_Transfer_entropy,Horowitz2014_Thermodynamics},
\begin{subequations}
    \begin{align}
        \dot \Sigma^X &:= \sum\limits_{x\ge x',y} J^{xx'}_{y} \ln \frac{R^{xx'}_{y}p_{x'y}}{R^{x'x}_{y}p_{xy}} \geq 0\\
        \dot \Sigma^Y &:= \sum\limits_{x, y\ge y'} J^{x}_{yy'} \ln \frac{R^{x}_{yy'}p_{xy'}}{R^{x}_{y'y}p_{xy}} \geq 0\,,
    \end{align}
\end{subequations}
expressing subsystem-specific second laws of thermodynamics. At steady state, the subsystem-specific entropy productions can be rewritten as
\begin{subequations}\label{eq:ep_info_flow}
\begin{align}
   0 \leq \dot\Sigma^X &= - \frac{\dot Q^X}{\kB T} + \dot I^Y \label{eq:ep_x_info_flow}\\
   0 \leq \dot\Sigma^Y &= - \frac{\dot Q^Y}{\kB T} - \dot I^Y\,, \label{eq:ep_y_info_flow}
\end{align}\label{eq:2nd_law}
\end{subequations}
in terms of the heat flows~\eqref{eq:heat_flows} and an \emph{information flow}~\cite{Ehrich2022_Energy} 
\begin{align}
    \dot I^Y &:= \sum\limits_{x,y\ge y'} J^{x}_{yy'} \ln\frac{p_{x|y}}{p_{x|y'}}, \label{eq:info_flow}
\end{align} 
for conditional probability $p_{x|y}$. The information flow quantifies how the dynamics of a given subsystem affect the mutual information between the two subsystems. If $\dot I^Y>0$, the $Y$ subsystem increases mutual information while $X$ consumes it, and $\dot I^Y < 0$ indicates that $Y$ decreases mutual information while $X$ increases it. Combining the fist laws in Eqs.~\eqref{eq:first_law} and the second laws in Eqs.~\eqref{eq:ep_info_flow} reveals that the input and output powers are each simultaneously bounded by an intermediate quantity~\cite{Barato2017_Thermodynamic_cost} called \emph{transduced capacity}~\footnote{Note that we here use units of energy, i.e., our transduced capacity is $\kB T$ times the definition in Refs.~\cite{Lathouwers2022_Internal,Ehrich2022_Energy,Leighton2022_Inferring}.} in Ref.~\cite{Lathouwers2022_Internal}:
\begin{align}
    \dot W^Y \geq \underbrace{\dot W^{Y \to X} + \kB T \dot I^Y}_{\mathrm{transduced}\,\mathrm{capacity}} \geq - \dot W^X\,. \label{eq:transduced_cap}
\end{align}
This equation illustrates that there are two pathways for free-energy transduction: the conventional transduced power $\dot W^{Y \to X}$ as well as information flow $\dot I^Y$. While the former has a straightforward interpretation in terms of the rate of work done by the $Y$-subsystem on the $X$-subsystem, the information flow is more subtle. In the extreme case with $\dot W^{Y \to X}=0$ and $\dot I^Y>0$, no work is transduced and the $X$-subsystem rectifies fluctuations using information generated by the $Y$-subsystem, a setup that can reasonably be described as a Maxwell demon or information engine~\cite{Leff2003_Maxwells,Parrondo2015_Thermodynamics,Horowitz2014_Thermodynamics,Hartich2014_Transfer_entropy,Freitas2021_Characterizing,Saha2021_Maximizing,Amano2022_Insights,Ehrich2022_Energy}. In the following we investigate which combination of these pathways maximizes the performance of our work-transducer model.

\section{Performance optimization}\label{performance optimization}
We assume the chemical-potential differences $\DmX$ and $\DmY$ driving the system are fixed by external constraints and optimize the parameters of the energy matrix $\epsilon_{xy}$ to maximize steady-state output power $-\dot W^X$.

\subsection{A single cycle maximizes output power}
\begin{figure}[tb]
\centering
\includegraphics[width=1\linewidth]{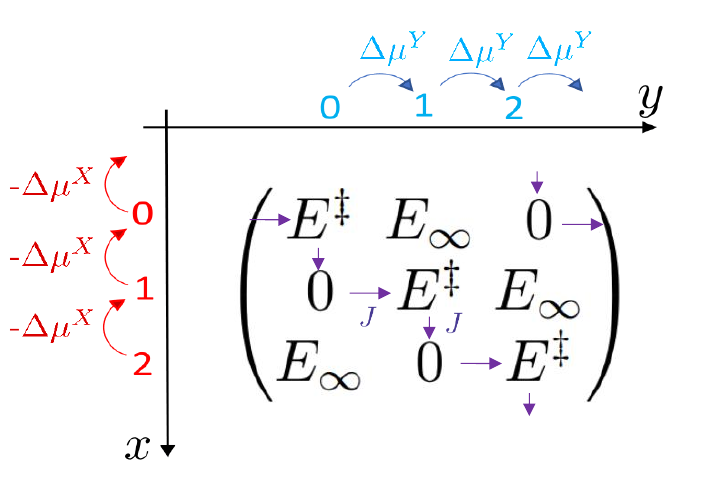}
\caption{Optimal energy matrix $\epsilon_{xy}$ and path of probability flux (purple arrows) when $E_\infty\rightarrow\infty$ and $\Delta\mu_Y>-\Delta\mu_X>0$. Blue: states of $Y$; red: states of $X$. Blue and red arrows indicate the directions of the respective driving forces on $Y$ and $X$.}
\label{fig:trajectory_infinite_a}
\end{figure}

In App.~\ref{app:gradient_descent} we use a gradient-descent algorithm to optimize the 9 entries $\epsilon_{xy}$ of the energy matrix. The resulting optimal energy matrix is intuitive: it produces a single cycle through state space that tightly couples input and output power. We therefore parameterize the matrix as 
\begin{align}
    \epsilon_{xy} = \begin{pmatrix}
        E^{\ddagger} & E_\infty & 0\\
        0 & E^{\ddagger} & E_\infty\\
        E_\infty & 0 & E^{\ddagger}
    \end{pmatrix}\,, \label{eq:coupling_matrix}
\end{align}
where $E^{\ddagger}$ is a finite energy [which we will exactly calculate later, see Eq.~\eqref{eq:c_opt_equal_temperatures}], and $E_\infty$ tends to infinity, forming an insurmountable energy barrier that channels the probability flux of the joint system along a single path through state space. The direction of this path is set by the magnitudes and directions of the driving forces $\Delta\mu^X$ and $\Delta\mu^Y$. Figure~\ref{fig:trajectory_infinite_a} illustrates the resulting probability flux.

Due to the tight coupling, in each full rotation $3\Delta\mu^X$ and $3\Delta\mu^Y$ flow between the machine and the respective chemical reservoirs. Hence the power ratio $\eta$~\eqref{eq:efficiency} is entirely determined by the ratio of the chemical driving forces, 
\begin{align}
    \eta = \frac{-\Delta\mu^X}{\Delta\mu^Y}\,, \label{efficiency steady state}
\end{align}
and notably is independent of the remaining free parameter $E^\ddagger$ in the energy matrix. Thus, to further optimize the motor at fixed $\Delta\mu^X$ and $\Delta\mu^Y$, we focus on optimizing the output power by varying the parameter $E^{\ddagger}$.

\subsection{Output power is maximized at vanishing information flow}\label{vanishing information}
\begin{figure}[htb]
  \centering 
  \includegraphics[width=\linewidth]{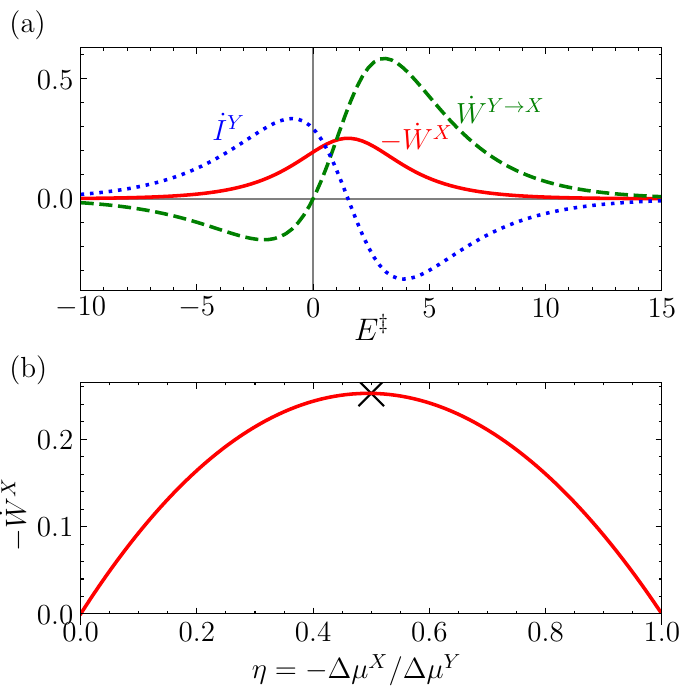}\\
  \caption{Performance optimization of the work transducer. (a) Output power $-\dot W^X$~\eqref{eq:work_flows_x} (red solid curve), transduced power $\dot  W^{Y \to X}$~\eqref{eq:internal_energy_flow_y} (green dashed curve), and information flow $\dot I^Y$~\eqref{eq:info_flow} (blue dotted curve), each as a function of the coupling parameter $E^\ddagger$ in the energy matrix~\eqref{eq:coupling_matrix}, for fixed driving strengths $\DmX = -1$ and $\DmY = 2$. (b) Maximal output power as a function of the power ratio, by using $E^\ddagger=E^{\ddagger}_\mathrm{opt}$ \eqref{eq:c_opt_equal_temperatures} and varying driving strength $\DmX$ at fixed $\DmY=2$. Black cross: $\DmX=-1$ used in (a).}
\label{fig:performance_optimization}
\end{figure}

Due to conservation of probability and the symmetry of the energy matrix, the probability fluxes $J^{xx'}_y$, $J^x_{yy'}$ along the path in Fig.~\ref{fig:trajectory_infinite_a} between adjacent states is uniform, i.e., $J^{x=y+1}_{y+1,y} = J^{x+1,x}_{y=x} =: J$. Explicitly,
\begin{subequations}\label{eq:flux_exp}
\begin{align}
    J &= R^{x=y+1}_{y+1,y}\, p_{x=y+1,y} - R^{x=y+1}_{y,y+1}\, p_{x=y+1,y+1} \label{eq:flux_y}\\
    &= e^{(\DmY-E^{\ddagger})/2}\, p_0 - e^{-(\DmY-E^{\ddagger})/2}\, p_{E^{\ddagger}} \label{eq:flux_y_expl}\\
    &= R^{x+1,x}_{y=x}\, p_{x,y=x} - R^{x,x+1}_{y=x}\, p_{x+1,y=x} \label{eq:flux_x}\\
    &= e^{(\DmX+E^{\ddagger})/2}\, p_{E^{\ddagger}} - e^{-(\DmX+E^{\ddagger})/2}\, p_0\,, \label{eq:flux_x_expl}
\end{align}
\end{subequations}
where we have used the rates in Eqs.~\eqref{eq:transition_rates_x} and \eqref{eq:transition_rates_y} and identified the steady-state probabilities $p_{E^{\ddagger}}:= p_{x=y,y}$ and $p_0 := p_{x=y+1,y}$ associated with the states of energy $E^{\ddagger}$ and $0$, respectively. Solving Eqs.~\eqref{eq:flux_y_expl} and \eqref{eq:flux_x_expl} and $p_{E^{\ddagger}} + p_0 = 1/3$ for $J$, $p_{E^\ddagger}$, and $p_0$ yields
\begin{subequations}
\begin{align}  
    J &= \frac{1}{3} \, \frac{\sinh{\frac{1}{2}\left(\DmX + \DmY\right)}}{ \cosh{ \frac{1}{2}\left(\DmX + E^{\ddagger} \right) } + \cosh{ \frac{1}{2}\left(\DmY - E^{\ddagger} \right)} } \label{eq:flux_sol}\\
    p_{E^{\ddagger}} &= { \frac{1}{6} \, \frac{e^{-(\DmX+E^{\ddagger})/2} + e^{(\DmY - E^{\ddagger})/2}}{ \cosh{ \frac{1}{2}\left(\DmX + E^{\ddagger} \right)} + \cosh{ \frac{1}{2}\left(\DmY - E^{\ddagger} \right)} } }\label{eq:pc_sol}\\
    p_0 &= {\frac{1}{6} \, \frac{{e^{(\DmX+E^{\ddagger})/2}} + e^{-(\DmY - E^{\ddagger})/2}}{ \cosh{ \frac{1}{2}\left(\DmX + E^{\ddagger} \right)} + \cosh{ \frac{1}{2}\left(\DmY - E^{\ddagger} \right)}} }\,.\label{eq:p0_sol}
\end{align}
\end{subequations}
This simplifies the output power $\dot W^X$ ~\eqref{eq:work_flows_x}, transduced power $\dot W^{Y \to X}$ \eqref{eq:internal_energy_flow_y}, and information flow $\dot I^Y$ \eqref{eq:info_flow} to
\begin{subequations}
\begin{align}
    \dot W^X &= 3 \, J \, \DmX \label{eq:single_path_output_work}\\
    \dot W^{Y \to X} &= 3\, J \, E^{\ddagger} \label{eq:single_path_transduced_work} \\
    \dot I^Y &=  3\, J \, \ln\frac{p_{E^{\ddagger}}}{p_0}\,,\label{eq:single_path_info_flow}
\end{align}\label{eq:all_thermo_quantities}%
\end{subequations}%
where in the last line we used that the marginal probabilities are uniform due to symmetry, i.e., $p_{y=0} = p_{y=1} = p_{y=2} =1/3$ and hence $p_{x=1|y=1}/p_{x=1|y=0} = p_{x=1,y=1}/p_{x=1,y=0} = p_{E^\ddagger}/p_0$, and similarly for the other two $Y$-transitions. Figure~\ref{fig:performance_optimization}(a) shows these quantities for $\DmX=-1$ and $\DmY=2$.

The value of $E^{\ddagger}$ controls how free energy is transduced in the machine. At $E^{\ddagger}=0$, no energy is transduced, hence the joint system acts like a pure information engine. At an intermediate value $E^{\ddagger}>0$, no information is transduced and the system acts as a conventional energy transducer. At any other $E^\ddagger$, both energy and information are transduced, rendering the system a hybrid engine.

Maximizing the flux $J$~\eqref{eq:flux_sol} over $E^{\ddagger}$ yields
\begin{align}
    E^{\ddagger}_{\rm opt} = \frac{\Delta\mu^Y-\Delta\mu^X}{2}\,, \label{eq:c_opt_equal_temperatures}
\end{align}
which averages the driving forces to produce equal forward rates along the single pathway,
\begin{align}
    R^{x=y+1}_{y+1,y}\Big|_{E^{\ddagger}_\mathrm{opt}} = R^{x+1,x}_{y=x}\Big|_{E^{\ddagger}_\mathrm{opt}} = e^{(\DmX+\DmY)/2}\,,
\end{align}
and similarly for the reverse rates. 

Figure~\ref{fig:performance_optimization}(b) shows the maximal output power as a function of the power ratio, obtained by substituting $E^{\ddagger}_\mathrm{opt}$~\eqref{eq:c_opt_equal_temperatures} in the energy matrix~\eqref{eq:coupling_matrix} and varying the chemical potential $\DmX$ at fixed $\DmY$. Maximum power is obtained at intermediate power ratio, and maximum power ratio is only possible at vanishing power when the engine stalls.

One might have naively expected $E^{\ddagger}=0$ to optimize the current by eliminating energy barriers; however, $E^{\ddagger}_\mathrm{opt}$ smooths the driving forces and ensures that no single transition is rate-limiting for the machine's progress through its state space. Interestingly, this choice of $E^{\ddagger}$ eliminates the information flow: With $E^{\ddagger}=E^{\ddagger}_\mathrm{opt}$, $p_{E^{\ddagger}} = p_0$ in Eqs.~\eqref{eq:pc_sol} and \eqref{eq:p0_sol}. Then, Eq.~\eqref{eq:single_path_info_flow} reads
\begin{align}
    \dot I^Y\big|_{E^{\ddagger}_\mathrm{opt}} = 0\,.\label{eq:no_information_flow}
\end{align}
We therefore conclude that performance is optimized when the engine behaves in a conventional way by transducing power and no information between its components, and that rectifying thermal fluctuations using an information-engine mechanism cannot provide as much output power. This finding echoes Ref.~\cite{Saha2022_Information}, where it was found that an information ratchet is less efficient at transporting a particle uphill than conventional ``dragging"; however, it was also found that when there are sufficient additional, nonthermal fluctuations, the information ratchet can be more efficient. In the following section we seek to approximate this effect by subjecting the different machine components to different temperatures, thereby creating an imbalance in thermal fluctuations.

In App.~\ref{app:splitting_factors} we demonstrate that, with different splitting factors for the $X$ and $Y$ dynamics, output power can be maximized at finite information flow. However, the power-maximizing information flow is small.

\section{Introducing different temperatures}
ATP synthase is located in a membrane at the interface of two different media. The $\mathrm{F}_{\rm o}$ subunit is embedded in a lipid membrane, while the $\mathrm{F}_1$ subunit is in aqueous surroundings~\cite{Boyer1997_ATP_Synthase}. It is therefore conceivable that the properties of the noises acting on each subsystem differ. Most likely the individual components are not only subjected to thermal fluctuations but also to additional nonequilibrium ones~\cite{Mizuno2007_Nonequilibrium,Gallet2009_Power}, although a difference in temperatures (albeit small relative to those found in classical examples of macroscopic heat engines) on each side of the membrane is also possible~\cite{Pinol2020_Real-Time,Wu2022_Intracellular,Di2022_Spatiotemporally}. To simplify the analysis, we here assume that subsystem $X$ is coupled to a heat bath at temperature $T^X$ and subsystem $Y$ is coupled to a heat bath at temperature $T^Y$, thereby subjecting the joint system to distinct noises of different strengths.

\subsection{Transmitted and received capacity}
When the subsystems are in contact with heat baths at different temperatures, the second laws~\eqref{eq:2nd_law} read 
\begin{subequations}
\begin{align}
   0 \leq \dot\Sigma^X &= - \frac{\dot Q^X}{\kB T^X} + \dot I^Y\\
   0 \leq \dot\Sigma^Y &= - \frac{\dot Q^Y}{\kB T^Y} - \dot I^Y\,.
\end{align}
\end{subequations}

Now, using the first laws~\eqref{eq:first_law} and replacing the temperature $T$ in Eq.~\eqref{eq:transduced_cap} with the respective temperatures $T^X$ and $T^Y$, we define different \emph{transmitted} and \emph{received} capacities:
\begin{subequations}
\begin{align}
    \dot W^Y &\geq \underbrace{\dot W^{Y\to X} + \kB T^Y \dot I^Y}_{\mathrm{transmitted}\,\mathrm{capacity}}\\
    -\dot W^X &\leq \underbrace{\dot W^{Y\to X} + \kB T^X \dot I^Y}_{\mathrm{received}\,\mathrm{capacity}}\,.
\end{align}
\end{subequations}
Here, in contrast to Eq.~\eqref{eq:transduced_cap}, input and output power are no longer simultaneously bounded by a single transduced capacity. The two righthand sides represent different rates of transduced free energy~\cite{Large2021_Free-energy} that bound the input power $\dot W^Y$ and output power $-\dot W^X$.

This illustrates the effect of a temperature difference: If $T^X > T^Y$ and $\dot I^Y > 0$, the received capacity is larger than the transmitted capacity and---relative to the equal-temperature case---more output power can be generated at the same input power. All else being equal, an increase in the energy flow produces the same increase in the minimum input power and in the maximum output power. However, the effect of a change in information flow is scaled by the temperature. When $T^X>T^Y$, a positive information flow $\dot I^Y$ can generate more output power in subsystem $X$ than is required for its generation in $Y$. On the other hand, when $T^X<T^Y$ with $\dot I^Y < 0$, more power is needed to generate information flow in subsystem $Y$ than can be harnessed in subsystem $X$. We therefore expect an optimized machine to utilize an information flow from cold to hot to maximize free-energy transduction.

With $T^X \neq 1$ and $T^Y \neq 1$, Eqs.~\eqref{eq:flux_exp} become~\footnote{For simplicity we assumed that the temperature does not affect the kinetic prefactor. Following Kramers theory~\cite{VanKampen2007}, larger temperatures also increase the bare transition rate, which would scale the resulting powers and information flows accordingly.}
\begin{subequations}
\begin{align}
    J &= e^{(\DmY-E^{\ddagger})/2 T^Y}\, p_0 - e^{-(\DmY-E^{\ddagger})/2T^Y}\, p_{E^{\ddagger}} \label{eq:flux_y_expl_tmp}\\
    &= e^{(\DmX+E^{\ddagger})/2T^X}\, p_{E^{\ddagger}} - e^{-(\DmX+E^{\ddagger})/2T^X}\, p_0\,, \label{eq:flux_x_expl_tmp}
\end{align}
\end{subequations}
and solving Eqs.~\eqref{eq:flux_y_expl_tmp} and \eqref{eq:flux_x_expl_tmp} together with $p_0 + p_{E^{\ddagger}} = 1/3$ yields
\begin{subequations}
\begin{align}
    J &= \frac{1}{3}\, \frac{\sinh \left(\frac{\DmY-E^{\ddagger}}{2T^Y} + \frac{\DmX+E^{\ddagger}}{2T^X} \right)}{\cosh \frac{\DmX+E^{\ddagger}}{2T^X} + \cosh \frac{\DmY-E^{\ddagger}}{2T^Y}}\\
    p_{E^{\ddagger}} &= \frac{1}{3}\,\frac{\exp\left(\frac{\DmY-E^{\ddagger}}{2T^Y}\right)}{\exp\left(\frac{\DmX+E^{\ddagger}}{2T^X}\right) + \exp\left(\frac{\DmY-E^{\ddagger}}{2T^Y}\right)}\\
    p_0 &= \frac{1}{3}\,\frac{\exp\left(\frac{\DmX+E^{\ddagger}}{2T^X}\right)}{\exp\left(\frac{\DmX+E^{\ddagger}}{2T^X}\right) + \exp\left(\frac{\DmY-E^{\ddagger}}{2T^Y}\right)}\,.
\end{align}
\end{subequations}
Substituting these into Eqs.~\eqref{eq:all_thermo_quantities} gives the output power $-\dot W$, transduced power $\dot W^{Y \to X}$, and information flow $\dot I^Y$.

\begin{figure}[t]
    \centering
    \includegraphics[width = \linewidth]{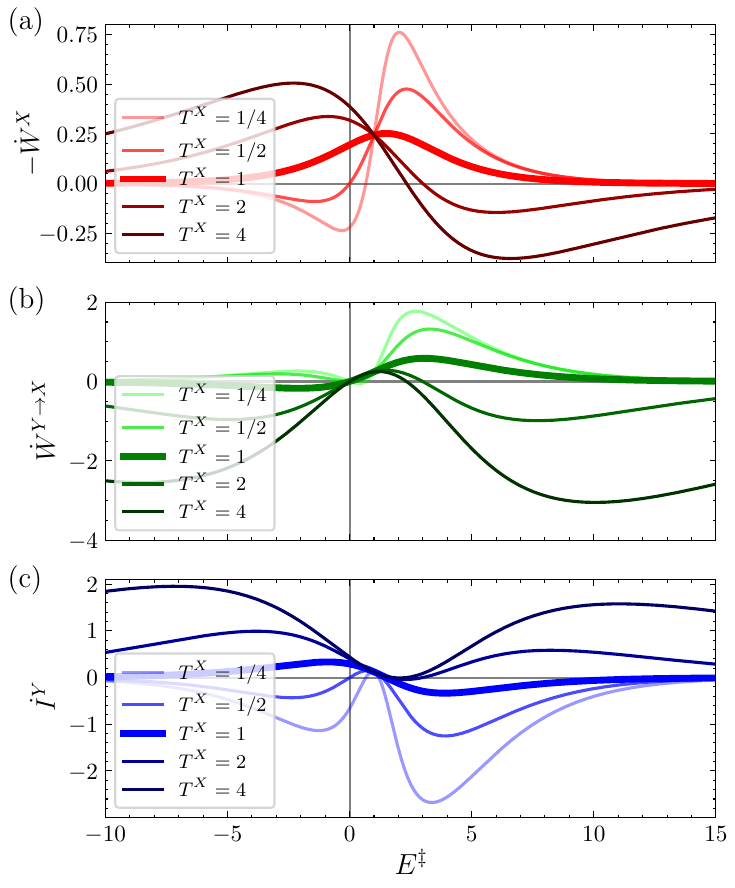}
    \caption{Work-transducer performance as a function of coupling parameter $E^{\ddagger}$, for different temperatures $T^X$ acting on the $X$ subsystem, at fixed $T^Y=1$, $\DmX=-1$, and $\DmY=2$, as in Fig.~\ref{fig:performance_optimization}(a). (a) Output power $-\dot W^X$. (b) Transduced power $\dot W^{Y \to X}$. (c) Information flow $\dot I^Y$. Darker colors indicate higher temperatures $T^X$. The plots for $T^X=1$ from Fig.~\ref{fig:performance_optimization}(a) are reproduced here with thicker curves.} 
    \label{fig:vary_temp_c}
\end{figure}

Figure~\ref{fig:vary_temp_c}(a) depicts the output power as a function of the coupling parameter $E^{\ddagger}$, for different temperatures and at fixed driving strengths $\DmX$ and $\DmY$. Having access to different temperature reservoirs allows for a larger output power when the coupling parameter is tuned appropriately. Figure~\ref{fig:vary_temp_c}(b) plots the transduced power $\dot W^{Y \to X}$. Generally, maximum output power is achieved when energy flows from hot to cold (from $X$ to $Y$ when $T^X>T^Y$ and from $Y$ to $X$ when $T^X < T^Y$). Transduced power is accompanied by information flow $\dot I^Y$, shown in Fig.~\ref{fig:vary_temp_c}(c). At maximum output power this information flow opposes the transduced power, i.e. information flows from $Y$ to $X$ when $T^X> T^Y$ and from $X$ to $Y$ when $T^X < T^Y$.

\begin{figure}[t]
    \centering
    \includegraphics[width = \linewidth]{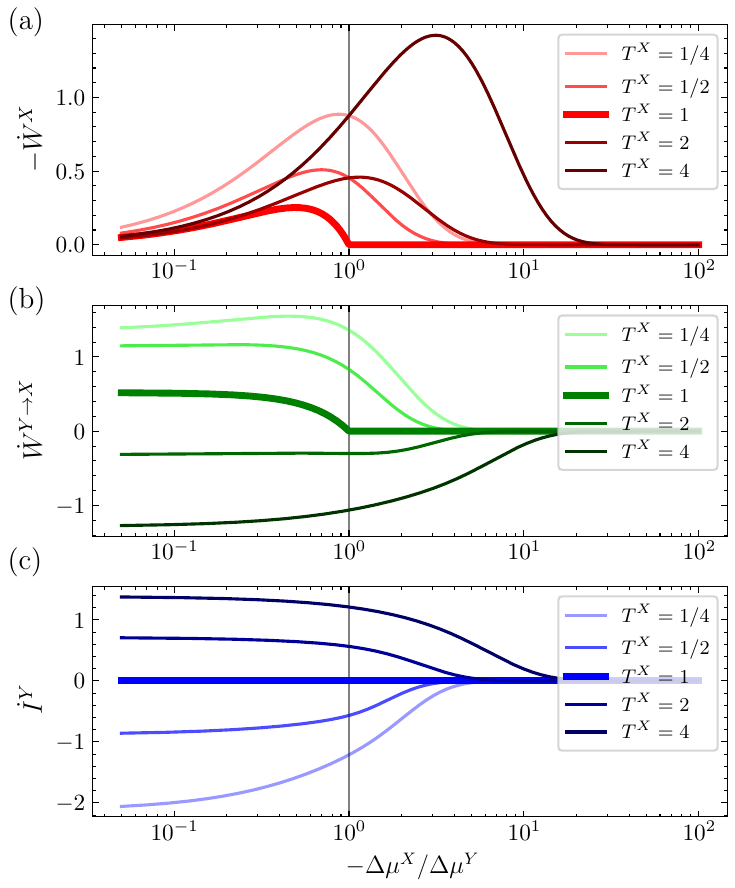}
    \caption{Performance at maximal output power, as a function of power ratio, at different temperatures $T^X$. (a) Output power maximized over coupling parameter $E^{\ddagger}$, for different temperatures $T^X$. (b) Transduced power $\dot W^{Y\to X}$. (c) Information flow. Here, $T^Y=1$ and $\DmY=2$, the same as for the corresponding equal-temperature plot in Fig.~\ref{fig:performance_optimization}(b). A cusp appears in (a) and (b) when $T^X=1$, because at equal temperatures output power cannot surpass input power and hence the optimized engine stalls for $-\DmX>\DmY$.}
    \label{fig:vary_temp_c_opt}
\end{figure}

Numerically maximizing the output power over the coupling parameter $E^{\ddagger}$ gives the equivalent of Fig.~\ref{fig:performance_optimization}(b), i.e., the trade-off between output power $-\dot W^X$ and power ratio $-\DmX/\DmY$, shown in Figure~\ref{fig:vary_temp_c_opt}(a). When the engine can exploit different temperatures, the maximal output power is greater. Additionally, the output power can exceed the input power. This is only possible because it is not only chemically driven but also allowed to exploit a temperature difference that produces an internal energy flow from hot to cold, as seen in Fig.~\ref{fig:vary_temp_c_opt}(b), and an information flow from cold to hot, shown in Fig.~\ref{fig:vary_temp_c_opt}(c).

\section{Discussion and Conclusion}
In this paper we have analyzed a simplified, discrete-state-space model inspired by the molecular machine $\mathrm{F}_\mathrm{o}\!-\!\mathrm{F}_1$ ATP synthase. To maximize the motor's output power, the coupling between the two motor components first needs to provide barriers to channel probability flux in such a way as to minimize slippage between the components. Moreover, the energy landscape should ``smooth over'' the forces generated by the chemical-potential differences that drive the rotation, to provide a constant net force driving the motor through its state space.

A machine optimized in this way transduces free energy by trading energy between its components. However, when relaxing the assumption of isothermality and allowing the different components to be in contact with heat baths at different temperatures, a second, entropic pathway to free-energy transduction becomes relevant. This entropic interaction is characterized by the information flow which describes how the dynamics of each subsystem modify the joint system's entropy balance. Information flow becomes relevant in non-isothermal settings because the minimum energetic cost to generate it and its maximum energetic benefit are respectively scaled by the ``input'' and ``output'' temperatures, making it worthwhile to exploit this pathway when those temperatures differ. By contrast, in isothermal settings it is more economical to directly transduce energy through the machine and forego any information transduction.

Strikingly, a temperature difference between the machine's components allows for output power larger than input power, suggesting that the machine exploits the temperature difference in the manner of a heat engine. This behavior is accompanied by an information flow. Naively, when the parameters are optimized to maximize output power, one might have expected both information flow and transduced power to always flow from the input to the output machine component. However, the direction of these flows correlates with the temperature difference: Power is transduced from hot to cold and information flows from cold to hot.

Our model uses just three states per subsystem; however, having more subsystem states could produce different optimization results. For instance, if the two motor components have varying state numbers, their ``gearing ratio'' can be adjusted~\cite{Foster1982_Stoichiometry} to match the relative strength of the chemical forces, while avoiding slippage between them.

Our model is a simplification of previous models of $\mathrm{F}_{\rm o}\!-\!\mathrm{F}_1$ ATP synthase that have a continuous state space~\cite{Lathouwers2020_Nonequilibrium,Lathouwers2022_Internal}. For these it was shown that the slippage between the components is important to maximize output power~\cite{Lathouwers2020_Nonequilibrium}, whereas our model has maximum output power at tight coupling. This is likely because the discrete model does not capture the effect of energy barriers between states, which are here accounted for only implicitly in the transition rates.

It would be interesting to analyze information flows in real-world molecular machinery, especially in $\mathrm{F}_\mathrm{o}\!-\!\mathrm{F}_1$ ATP synthase~\cite{Yasuda2001,Toyabe2010_Nonequilibrium,Toyabe2011_Thermodynamic,Kawaguchi2014_Nonequilibrium,HayashiPRL15}. While subsystem efficiencies can be estimated from the average behavior of the joint system~\cite{Leighton2022_Inferring}, estimating information flows necessitates observing the detailed dynamics of both subunits. Synthetic chemical motors~\cite{Penocchio2022_Information,Amano2022_Insights} and the dimeric stepping motors kinesin and myosin~\cite{Takaki2022_Information} have recently been analyzed in this way. Our model system suggests that information flow plays a minor role in free-energy transduction when fluctuations on both subsystems are comparable. However, when the fluctuations are different (e.g., because of a temperature difference), information flow can become a valuable resource.

\begin{acknowledgements}
We thank Matthew P.\ Leighton for insightful discussions and feedback on the manuscript and Zak Frentz for sharing helpful insights into cellular temperature differences. This research was supported by a Mitacs Globalink Research Internship (M.G.); the Foundational Questions Institute, a donor-advised fund of the Silicon Valley Community Foundation, grant FQXi-IAF19-02 (D.A.S. and J.E.); a Natural Sciences and Engineering Research Council of Canada (NSERC) Discovery Grant (D.A.S.); and a Tier-II Canada Research Chair (D.A.S.).
\end{acknowledgements}

\appendix
\section{Gradient-descent optimization of energy landscape} \label{app:gradient_descent}
\begin{figure}[htb]
    \centering
    \includegraphics[width = \linewidth]{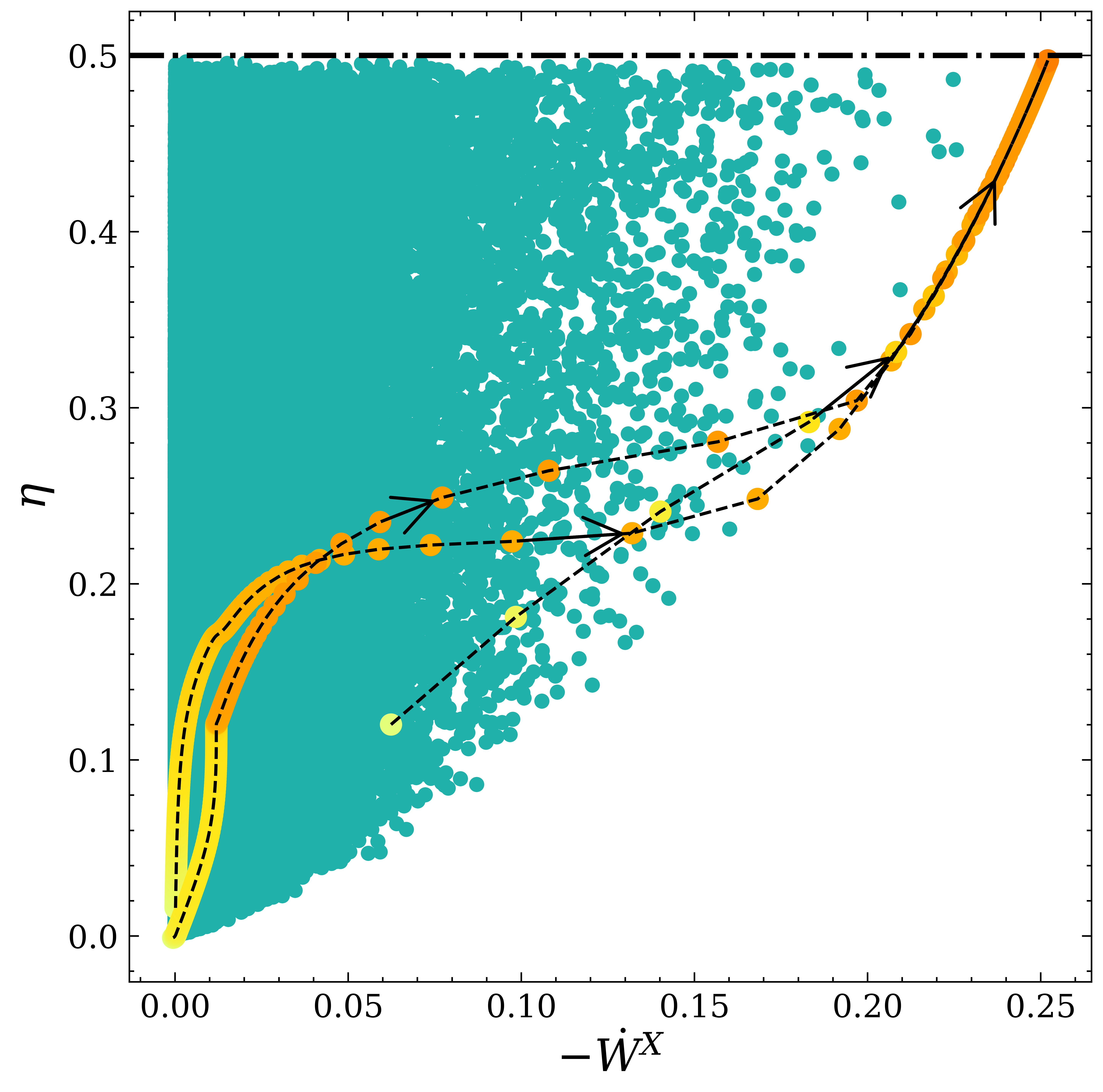}
    \caption{Parametric plot of output power $-\dot W^X$~\eqref{eq:work_flows_x} and power ratio $\eta$~\eqref{eq:efficiency} for $\DmX = -1$ and $\DmY = 2$. Teal dots show $10^7$ random energy matrices $\epsilon_{xy}$ with coefficients uniformly sampled between 0 and 20. Yellow-to-orange points connected by dashed black lines depict three trajectories of the gradient descend algorithm started from random initial conditions with arrows indicating trajectory direction. The dash-dotted line indicates the power ratio at tight coupling~\eqref{efficiency steady state}.}
    \label{fig:gradient_plot} 
\end{figure}

We vary the nine coefficients in the energy matrix $\epsilon_{xy}$ to maximize the output power $-\dot W^X$~\eqref{eq:work_flows_x} and power ratio $\eta$~\eqref{eq:efficiency}. Generating random matrices suggests that output power and power ratio can be simultaneously maximized (see scatterplot in Fig.~\ref{fig:gradient_plot}).

To find the optimal coefficients, we employ a gradient-descent algorithm~\cite{Ruder2016AnOO}. We start with random coefficients $\epsilon_{xy}$ and vary them according to
\begin{align}\label{eq: gradient_descent2}
    \epsilon_{xy}^{k+1}=\epsilon_{xy}^{k}+\alpha_{1}\frac{\partial \eta}{\partial \epsilon_{xy}}-\alpha_{2}\frac{\partial \dot W^X}{\partial \epsilon_{xy}}\,,
\end{align}
with positive \emph{learning rates} $\alpha_1$ and $\alpha_2$, and $k$ enumerating the iteration of the algorithm. The trajectory followed by this optimization is plotted in Fig.~\ref{fig:gradient_plot} and terminates at a ``tight-coupling'' energy matrix that allows only a single path through state space [Eq.~\eqref{eq:coupling_matrix}]. To avoid numerical divergences, we limited the coefficient range to [0,20]. The fact that no random parameter set outperformed the gradient-descent algorithm gives us confidence that we have found the global optimum. The apparent maximum $\eta=-\DmX/\DmY$ corresponds to the ``tight-coupling'' limit~\eqref{efficiency steady state} in which output power and power ratio can be maximized simultaneously.

\begin{widetext}
\section{Bipartite jump dynamics with different splitting factors} \label{app:splitting_factors}
Using Eqs.~\eqref{eq:transition_rates_x} and \eqref{eq:transition_rates_y}, we assumed that the energy difference between adjacent states symmetrically affect the forward and backward transition rates. However, thermodynamics only constrains the ratio of transition rates. Here, we investigate how an asymmetric sensitivity of transition rates on the energy difference (also called the splitting factor~\cite{Brown2020_Theory}) affects the results. We rewrite Eqs.~\eqref{eq:transition_rates_x} and \eqref{eq:transition_rates_y} with arbitrary splitting factors $s^X$ and $s^Y$, to obtain
\begin{subequations}
\begin{align}
     R^{x=y+1}_{y+1,y} & = e^{\; s^Y (\DmY-E^{\ddagger})} \label{eq:transition_rates_y_splitting_factor1} \\
     R^{x=y+1}_{y,y+1} & = e^{\;-(1-s^Y) (\DmY-E^{\ddagger})} \label{eq:transition_rates_y_splitting_factor2}\\
     R^{x+1,x}_{y=x} & = e^{\; s^X (\DmX+E^{\ddagger})} \label{eq:transition_rates_x_splitting_factor1}\\
     R^{x,x+1}_{y=x} & = e^{\; -(1-s^X) (\DmX+E^{\ddagger})} \, . \label{eq:transition_rates_x_splitting_factor2}
\end{align}
\end{subequations}
Note that the transition-rate ratio still fulfills local detailed balance~\eqref{eq:generalized_det_balance}. Using $p_{x=y+1,y} = p_0$ and $p_{x=y,y}= p_{E^{\ddagger}}$ for all $y$, and $p_0 + p_{E^{\ddagger}} = 1/3$, we solve Eqs.~\eqref{eq:flux_y} and \eqref{eq:flux_x} with the above rates to obtain flux
\begin{align}
    J =\frac{1}{3} \frac{ e^{(\DmY-E^{\ddagger}) s^Y+(\DmX+E^{\ddagger}) s^X}-e^{(\DmY-E^{\ddagger})\left(s^Y-1 \right)+(\DmX+E^{\ddagger}) \left(s^X-1\right)}}{e^{(\DmY-E^{\ddagger})s^Y}+e^{(\DmX+E^{\ddagger}) s^X}+e^{\left(\DmY-E^{\ddagger}\right) \left(s^Y-1\right)}+e^{\left(\DmX+E^{\ddagger}\right) \left(s^X-1\right)}}\, ,\label{eq:flux_splitting_factor}
\end{align}
and steady-state probabilities
\begin{subequations}
\begin{align}
    p_{E^{\ddagger}} = \frac{1}{3} \frac{e^{(\DmY-E^{\ddagger})s^Y}+e^{(\DmX+E^{\ddagger})(s^X-1)}}{e^{(\DmY-E^{\ddagger})s^Y}+e^{(\DmX+E^{\ddagger}) s^X}+e^{\left(\DmY-E^{\ddagger}\right) \left(s^Y-1\right)}+e^{\left(\DmX+E^{\ddagger}\right) \left(s^X-1\right)}}\, , \\
    p_0 = \frac{1}{3} \frac{e^{(\DmY-E^{\ddagger})(s^Y-1)}+e^{(\DmX+E^{\ddagger})s^X}}{e^{(\DmY-E^{\ddagger})s^Y}+e^{(\DmX+E^{\ddagger}) s^X}+e^{\left(\DmY-E^{\ddagger}\right) \left(s^Y-1\right)}+e^{\left(\DmX+E^{\ddagger}\right) \left(s^X-1\right)}}\, . 
\end{align} 
\end{subequations}
Inserting the flux $J$ and the probabilities $p_0$ and $p_{E^{\ddagger}}$ into Eqs.~\eqref{eq:single_path_output_work},~\eqref{eq:single_path_transduced_work},~and \eqref{eq:single_path_info_flow} yields output power $-\dot W^X$, transduced power $\dot W^{Y \to X}$, and information flow $\dot I^Y$. Figure~\ref{fig:vary_splitting_factor_c} shows these as a function of the coupling parameter $E^{\ddagger}$, illustrating qualitatively similar curves as in Fig.~\ref{fig:performance_optimization}(a).

\begin{figure}[htb] 
    \centering
    \includegraphics[width = 0.8\linewidth]{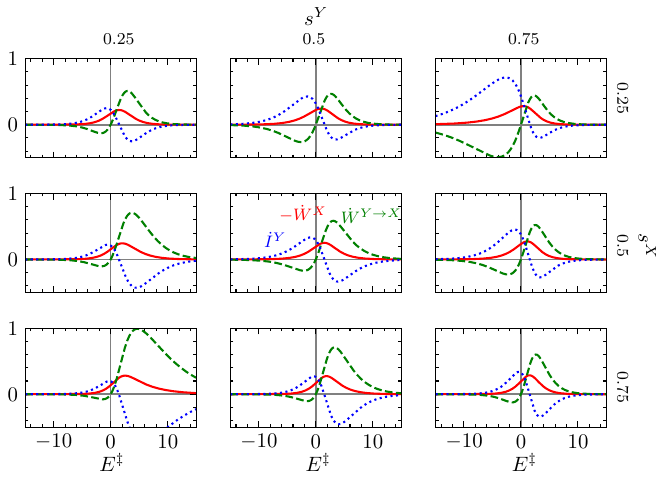}
    \caption{Work-transducer performance with different splitting factors $s^X$ and $s^Y$, at equal temperatures. Output power $-\dot W^X$~\eqref{eq:work_flows_x} (red solid curve), transduced power $\dot  W^{Y \to X}$~\eqref{eq:internal_energy_flow_y} (green dashed curve), and information flow $\dot I^Y$~\eqref{eq:info_flow} (blue dotted curve), each as a function of the coupling parameter $E^{\ddagger}$ in the energy matrix~\eqref{eq:coupling_matrix}, for fixed driving strengths $\DmX = -1$ and $\DmY = 2$. The middle plot corresponds to Fig.~\ref{fig:performance_optimization}(a).}
    \label{fig:vary_splitting_factor_c}
\end{figure}
\end{widetext}

\bibliography{references}
\end{document}